\documentclass[10pt]{iopart}
\usepackage{graphicx}

\usepackage{epsfig}
\begin{document}

\article[PhD Tutorial]{PhD Tutorial}{Cold collisions in dissipative optical lattices}

\author{J Piilo\dag\ and  K-A Suominen\ddag}

\address{\dag\ School of Pure and Applied Physics, University
of KwaZulu-Natal, Durban 4041, South Africa}

\address{\ddag\ Department of Physics, University of Turku, 
FIN-20014 Turun yliopisto, Finland}

\begin{abstract}
The invention of laser cooling methods for neutral atoms
allows optical and magnetic trapping of cold atomic clouds
in the temperature regime below $1$mK.
In the past, light-assisted cold collisions between 
laser cooled atoms have
been widely studied in magneto-optical atom traps (MOTs).
We describe here theoretical studies of dynamical interactions,
specifically cold collisions,
between atoms trapped in near-resonant, dissipative optical lattices.
The extension of collision studies to the regime of optical lattices
introduces several complicating factors.
For the lattice studies, one has to account
for the internal substates of atoms, position dependent
matter-light coupling, and position dependent
couplings between the atoms, in addition to the 
spontaneous decay of electronically excited atomic states.
The developed one-dimensional quantum-mechanical model combines atomic
cooling and collision dynamics in a single
framework. The model is based on Monte Carlo wave-function
simulations and is applied when the lattice-creating
lasers have frequencies both below (red-detuned lattice) and 
above (blue-detuned lattice) the atomic resonance frequency.
It turns out, that 
the radiative heating
mechanism affects the dynamics of atomic cloud in a
red-detuned lattice in a way that is not directly
expected from the MOT studies. 
The optical lattice and position dependent
light-matter coupling introduces selectivity
of collision partners.
The atoms, which are most
mobile and energetic, are strongly favored to participate in collisions,
and  are more often ejected from the lattice,
than the slow ones 
in the laser parameter region selected
for study. Consequently, the atoms remaining in the lattice
have a smaller average kinetic energy per atom than
in the case of non-interacting atoms.
For blue-detuned lattices, we study how optical shielding emerges as
a natural part of the lattice and look for ways to optimize the effect. 
We find that the cooling and shielding dynamics do not mix and it is
possible to achieve efficient shielding 
with a very simple arrangement. 
The simulations are computationally very demanding 
and would obviously benefit
from the simplification schemes.
We present some steps to this direction
by showing how it is possible to calculate
collision rates in near-resonant lattices
in a fairly simple way. The method can then be used
to combine quantum-mechanical and
semiclassical models for cold collision studies
in optical lattices.

\end{abstract}


\submitto{\JOB}

\maketitle


\section{Introduction}

It has been known for a long time that
light can exert a force on material 
objects. The first experiments
which showed the effect of the radiation pressure
of an electromagnetic field on matter predicted by
Maxwell were done at the beginning of the last century \cite{Nichols01a}.
Since then, the success of applying the light force in a controlled way
to cool gaseous atomic
clouds to temperatures around, and even below, the $\mu$K range 
has given a huge impetus to the field of cold atomic physics.

Laser cooling and trapping of neutral atoms has been a
very rapidly developing field of physics since 
the mid-eighties when experimentalists succeeded
for the first time to optically trap laser cooled atoms
\cite{Chu86a}. More recent milestones include
the experimental realizations
of atomic Bose-Einstein condensate (BEC) \cite{Anderson95a}, Fermi degenerate
dilute quantum gas \cite{DeMarco99a}, superfluid--Mott-insulator
phase transition in optical lattices \cite{Greiner02a},
molecular BEC \cite{Greiner03a}, and 
evidence 
for superfluidity in an atomic Fermi gas \cite{Chin04a}.

In general, the slow motion of the laser cooled atoms
changes drastically the collision dynamics of the atoms compared
to room temperature gases. The collisions may become
inelastic and 
the research on cold collisions in magneto-optical traps (MOTs)
has shown how the collisions consequently limit the densities and 
temperatures of atomic gases in MOTs \cite{Suominen96a,Weiner99a}.
The purpose of this
article is to describe and review the work
which has been done to extend the regime
of cold collision studies to optical lattices.

This tutorial gives a simple introduction to 
one specific cooling and trapping scheme for neutral
atoms: near-resonant, dissipative
optical lattices, and concentrates on the description
of light-assisted cold collisions in this system.
A thorough introduction into the field of
laser cooling, trapping, and optical lattices can
be found from a text book \cite{Metcalf99a}, various
review 
articles \cite{Stenholm86a,Jessen96a,Adams97a,Meacher98a,Rolston98a,
Guidoni99a,Grynberg01a,Bloch04a},
and summer school lecture notes
\cite{Cohen-Tannoudji92b,Grynberg96a,
Hemmerich96a}. Cold collision theories and experiments
are reviewed in detail in Refs.~\cite{Suominen96a,Weiner99a,Burnett95a}.

The article is organized as follows. Section \ref{sec:optlat} gives
a short introduction to near-resonant, dissipative optical lattices.
We describe both the red- and blue-detuned lattices.
The latter 
are sometimes in the literature referred also as gray or dark optical lattices
due to the reduced number of scattered photons
compared to "bright" red-detuned optical lattices.
Section \ref{sec:colcol} describes the basic cold collision mechanisms
in the presence of a near-resonant light, radiative
heating by the red-detuned light and optical
shielding by the blue-detuned light.
Section \ref{CollisionsInLattice}
shows a formulation of  the cold collision problem in
optical lattices and overviews the central results.
Finally, we conclude with a few remarks
in Section \ref{sec:con}. 

\section{Optical lattices}\label{sec:optlat}

A disordered cold atomic gas can be 
arranged into an ordered structure 
by periodic optical potentials
which are created with suitably chosen set of
interfering laser beams
\cite{Jessen96a,Meacher98a,Rolston98a,Guidoni99a,Grynberg01a,Bloch04a}.
An optical lattice formed
by optical potentials does not only trap
the atoms, but it is also able to cool them
via optical pumping and Sisyphus mechanism \cite{Dalibard89a,Ungar89a}.

The typical temperature range for alkali metal atoms
cooled and trapped in optical lattices
falls between the so called Doppler limit $T_d$ and
recoil limit $T_r$. The Doppler limit gives
lower bound for Doppler cooling,
the main cooling mechanism in MOTs, and
can be written as
\begin{equation}
T_d=\hbar \Gamma/2k_B,
\end{equation}
where $\hbar$ is Planck's constant $h$ divided by $2\pi$, $\Gamma$ the atomic linewidth,
and $k_B$ Boltzmann's constant.
 The recoil limit in turn describes the lower
bound for polarization gradient cooling mechanisms,
and $T_r$ corresponds to the amount of energy
of single photon absorption or emission recoil, and is given by
\begin{equation}
T_r=\left( \hbar k_r \right)^2 / M k_B.
\end{equation}
Here $k_r$ is the wavenumber of the cooling lasers, and 
$M$ the mass of an atom. The corresponding energy unit
\begin{equation}
E_r=\left( \hbar k_r \right)^2 /2 M,
\end{equation}
is consequently called recoil energy.
A typical alkali element used 
for laser cooling, Cs, has $T_d=120\mu$K and $T_r=0.2\mu$K.

The modern work on optical lattices was preceded by the
proposal of Letokhov to trap cold atoms in one dimension by
using a standing light wave
\cite{Letokhov68a}, and accompanied by the experiment of Burns {\it et al.}
who created crystal structures of microscopic dielectric objects suspended
in liquid by standing light waves \cite{Burns90a}. 
Because of the purity and ease of control
in optical lattices,
the trapped atoms 
are almost ideal
for the study of phenomena
which are familiar from solid-state physics. A few examples 
of these phenomena, which have been realized in experiments,
include the Bragg scattering \cite{Birkl95a,Weidemuller95a},
Bloch oscillations \cite{Bendahan96a}, and 
Wannier-Stark ladders \cite{Wilkinson96a}.
Other interesting observations include the quantum Zeno effect
\cite{Fischer01a} and dynamical tunneling \cite{Hensinger01a}.
During the last few years, the amount of research exploiting optical
lattices has hugely increased. This is mainly because
now it is possible to load 
far-off resonant optical lattices with Bose-Einstein condensates
(BECs). This has led to the observation
of superfluid--Mott-insulator phase transition
in far-detuned optical lattices \cite{Greiner02a}
which paved the way e.g. to the creation
of multiparticle entanglement by exploiting
controlled collisions \cite{Mandel03a}.
For dissipative optical lattices, a so called
double lattice creates interesting prospects
for the future studies
\cite{Ellmann03a}. In a double lattice, the atoms on 
two ground hyperfine states can be trapped and moved
in a controllable way with respect to each other.
This opens a new avenue to study inelastic cold collisions 
in the presence of near-resonant light in dissipative lattices.

\subsection{Red-detuned dissipative optical lattices}

\begin{figure}[tb]
\centering
\psfig{figure=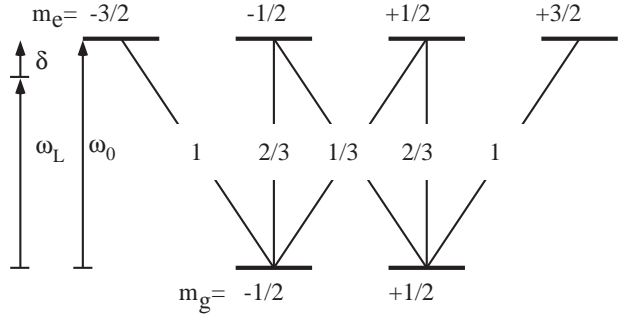,scale=0.4}
\caption[f2]{\label{fig:levels1}
The level structure of a single atom for
a red-detuned lattice. The squares of the Clebsch-Gordan
coefficients for various transitions are shown, and the laser frequency 
$\omega_L$ is detuned a few atomic linewidths
below the atomic resonance frequency $\omega_0$ for
efficient Sisyphus cooling. The detuning of the laser is described by
 $\delta=\omega_L-\omega_0$.
}
\end{figure}

The simplest case of cooling in
optical lattices can be described by using 
the atomic level structure with 
the ground state angular momentum
$J_g=1/2$ and the excited
state $J_e=3/2$ \cite{Dalibard89a}. 
A single atom has
two ground
state sublevels $|g_{\pm 1/2}>$ and four excited state sublevels
$|e_{\pm 3/2}>$
and $|e_{\pm 1/2}>$, where  the half--integer subscripts indicate the
quantum number
$m$ of the angular momentum along the $z$ direction.
This is shown in 
Fig.~\ref{fig:levels1} with the appropriate
squares of the Clebsch-Gordan
coefficients which describe the relative strengths of 
the light-induced couplings between the various levels.

\subsubsection{Sisyphus cooling}

\begin{figure}[tb]
\centering
\psfig{figure=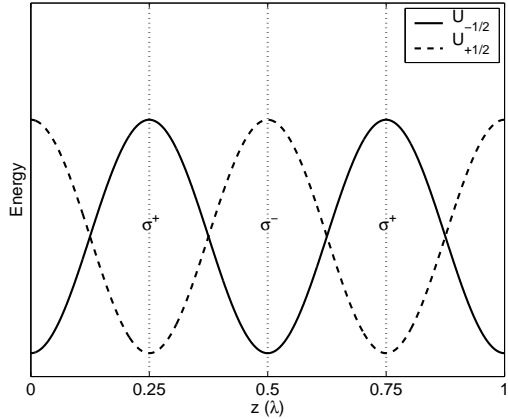,scale=0.4}
\caption[f2]{\label{fig:Lattice}
Schematic view of the optical potentials for the two ground
state Zeeman sublevels in a red-detuned lattice
with the atomic level structure $J_g=1/2, J_e=3/2$. The lattice structure is
created due to the periodic polarization gradient of the laser field, and
the points of pure circular polarization are indicated
by dotted lines.
}
\end{figure}

The laser field consists of two counter--propagating beams 
along the $z$-axis with
orthogonal linear
polarizations and with frequency $\omega_L$. The total field has a polarization
gradient in one dimension and reads 
(with suitable choices of phases
of the beams and origin of the coordinate system)
\begin{equation}
      {\bf E}(z,t)={\cal E}_0 ({\bf e}_x e^{ik_rz} - i {\bf e}_y
      e^{-ik_rz})e^{-i\omega_L t} + c.c.,
      \label{eq:Efield}
\end{equation}
where ${\cal E}_0$ is the amplitude and $k_r$ the wavenumber. With
this field, the
polarization changes from circular $\sigma^-$ to linear and back to
circular in the
opposite direction $\sigma^+$  when $z$ changes by $\lambda_L/4$
where $\lambda_L$
is the wavelength of the lasers.
The periodic polarization gradient of the laser field is reflected in
the periodic light shifts, i.e., AC--Stark shifts, 
of the atomic sublevels creating the optical
lattice structure. Figure \ref{fig:Lattice} 
displays a schematic view of the optical potentials and 
shows how the lattice wells coincide with the points of pure circular polarization of the light
field.

\begin{figure}[tb]
\centering
\psfig{figure=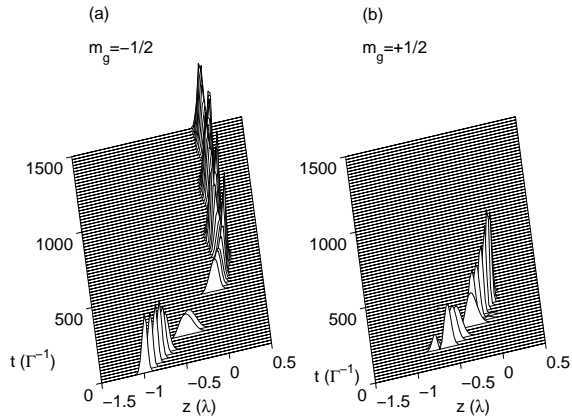,scale=0.4}
\caption[f3]{\label{fig:Localization}
Sisyphus cooling and the localization of an atom into the optical 
lattice. A
possible time evolution for a single atom wave packet is shown 
for two ground state
Zeeman levels: (a) $m_g=-1/2$ and (b) $m_g=+1/2$. 
The result shows the optical pumping cycles and the 
localization of a
single atom into the optical lattice. The discontinuous changes between the two 
ground states are
due to quantum jump events from the excited state (not shown), 
selected to happen
randomly with an appropriately weighted probability. If the run is 
repeated, the
jumps would appear at different times.}
\end{figure}

When the atomic motion occurs in a suitable velocity range, the optical
pumping of the atom between the ground state sublevels 
reduces the kinetic energy of the atom~\cite{Dalibard89a}.
After several cooling cycles the
atom localizes into an optical potential well, i.e., into an optical
lattice site.
Figure~\ref{fig:Localization} shows the optical pumping cycles
between the ground
state sublevels and the oscillations of the atomic wave packet
after localization into a lattice site.

The intensity of the laser field and the strength of the coupling
between the field
and the atom is described by  the Rabi frequency 
\begin{equation}
\Omega = 2 d {\cal
E}_0 / \hbar,
\end{equation}
where $d$ is  the atomic dipole moment of the strongest
transition between the ground and excited states. The detuning of the
laser field
from the atomic  resonance is given by 
\begin{equation}
\delta = \omega_L - \omega _{0},
\end{equation}
where $\omega_0$ is the atomic
resonance frequency.
As a unit for $\Omega$ and $\delta$ the atomic
linewidth $\Gamma$ is commonly used.

The Hamiltonian for a single atom moving in the  laser field
given in Eq.(\ref{eq:Efield}) is after the rotating wave approximation
\begin{equation}
     H =  \frac{p^{2}}{2M} - \hbar \delta
P_{e}
      + {V} \label{eq:Hred}.
\end{equation}
Here, $p^{2}/{2M}$ is the kinetic energy, $\delta$
the detuning of the laser,
$P_{e} =\sum_{m=-3/2}^{3/2} |e_m \rangle~ \langle
e_m|$ is the projection operator onto the excited state,
and the interaction between a single atom and the field is
\begin{eqnarray}
     {V}&=& -i\frac{\hbar\Omega}{\sqrt{2}} \sin(k_rz)
     \left\{|e_{3/2} \rangle\langle g_{1/2}| 
 + \frac{1}{\sqrt{3}} 
     |e_{1/2} \rangle\langle g_{-1/2}|\right\} \nonumber \\
     && +\frac{\hbar\Omega}{\sqrt{2}}\cos(k_rz)
     \left\{|e_{-3/2} \rangle\langle g_{-1/2}|
 + \frac{1}{\sqrt{3}} 
     |e_{-1/2} \rangle\langle g_{1/2}|\right\} +H.c.,
     \label{eq:VAtomLaser}
\end{eqnarray}
where $z$ is the position operator of the atom.
The spatially modulated optical potentials read
\begin{eqnarray}
     U_-=U_0\sin^2(k_rz), \nonumber \\
     U_+=U_0\cos^2(k_rz), \label{eq:U+-}
\end{eqnarray}
for ground states $m_g=-1/2$ and  $m_g=+1/2$,
respectively~\cite{Castin90a}. Here, $U_0$ is the modulation depth of the lattice.

\subsubsection{Localization in lattice}

When the steady state is reached after a certain period of cooling,
atoms are, to a
large extent, localized into the lattice sites.
The optical lattice is in the oscillating 
regime if the atom completes on average more
than one oscillation in the site before
being optically pumped
to a neighboring site.
For less than one average oscillation per site, the lattice
is in the jumping regime. 
The laser parameters
$\Omega$ and $\delta$ determine 
in which of the regimes the lattice is 
in~\cite{Castin90a}. 
It must be noted that tight localization and
occupation of
the lowest vibrational levels of a periodic lattice potential
increases the optical pumping
time, and the time of localization within a single lattice site becomes
longer compared to the semiclassical values \cite{Dalibard89a}.
Since we are mainly interested in the case when the two
atoms undergo an intra-well collision, the
chosen parameters in our work correspond
to the jumping regime of the lattice.

In our studies, the laser field is detuned a few
atomic linewidths below \cite{Piilo01a,Piilo02a,Piilo03a} or above 
\cite{Piilo02b} the atomic transition. The most typical 
detuning range we use is $3\Gamma\leq \delta\leq 10\Gamma$
towards the oscillating regime. 
Whereas the most common range in the dissipative
lattice  experiments done so far is slightly higher, $5\Gamma\leq \delta\leq 20\Gamma$, and in the oscillating regime.

\subsection{Blue-detuned dissipative optical lattices}

Two counterpropagating orthogonally polarized
blue-detuned ($\delta>0$) laser beams can efficiently cool down 
atoms which have the level structure 
$J_e=J_g$ or $J_e=J_g-1$~\cite{Grynberg01a}. The lowest
position-dependent eigenstate of the system
is flat in this case and is not directly coupled to the light
field at any point of space. 
Thus, despite of the cooling, the atoms
are not efficiently trapped.

The problem can be circumvented by the use of
either transverse \cite{Hemmerich95a} or longitudinal \cite{Grynberg94a}
magnetic fields with
respect to the laser propagation 
axis. We have used for
collision studies the proposal of Grynberg and Courtois with a magnetic
field along the laser axis \cite{Petsas96a}, see below. We note that
there exists also possibilities to create blue-detuned gray lattices
by all-optical means~\cite{Esslinger96a,Stecher97a}.

\subsubsection{Grynberg-Courtois gray optical
lattice}\label{sec:GC}

An applied longitudinal magnetic field removes the degeneracy
of the atomic states of blue-detuned optical
molasses mentioned above,
and produces the necessary
spatial modulation for the optical potentials
and the lattice structure~\cite{Grynberg01a}.

In general, the atomic states are now Zeeman shifted by
the magnetic field, and light shifted by the laser.
The ratio between the two shifts can be varied
by changing the intensity of the laser
or the strength of the magnetic field. 
Obviously, the two
extreme regimes are: a) the Zeeman 
shift is small compared to the light 
shift b) the light shift is small
compared to the Zeeman shift.

In case a), where the light shift dominates, the behavior
of the lattice is paramagnetic and the increasing magnetic field
strength increases the lattice modulation depth. 
When the Zeeman shift dominates in the case b), 
the laser field produces perturbations to the
Zeeman shifted states. The situation
resembles now the traditional Sisyphus cooling scheme
and the lattice is in the antiparamagnetic regime.

\begin{figure}[t!]
\noindent\centerline{\psfig{width=70mm,file=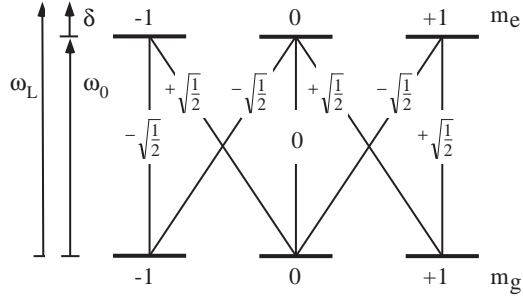} }
\caption[f1]{\label{fig:levels2}
The level structure of a single atom for a blue-detuned lattice
with the Clebsch-Gordan coefficients of
corresponding transitions. This structure corresponds to particular
hyperfine states of $^{87}$Rb and can be used for efficient
cooling and trapping of atoms in the
Grynberg-Courtois blue-detuned lattice in the antiparamagnetic
regime \cite{Petsas96a}.
}
\end{figure}

\begin{figure}[t!]
\noindent\centerline{\psfig{width=70mm,file=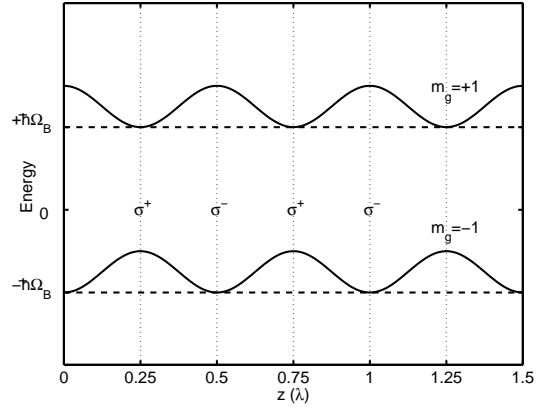} }
\caption[f2]{\label{fig:opot2}
Schematic view of the optical potentials for the two trapping ground
state Zeeman sublevels in a blue-detuned
lattice. The periodic polarization gradient of the laser
field creates the lattice structure and the points of circular
polarizations are indicated by $\sigma^{+}$ and $\sigma^{-}$. 
The dashed lines give the
Zeeman shifted energy levels which the light field modifies.}
\end{figure}

We have done our collision studies
in this antiparamagnetic regime. See Fig.~\ref{fig:levels2} for the used
atomic level configuration $J_e=J_g=1$ and Fig.~\ref{fig:opot2}
for the corresponding optical lattice structure.
We label the three ground state sublevels with
$|g_{\pm 1}>,|g_{0}>$, and the three excited state sublevels with $|e_{\pm 1}>,
|e_{0}>$, where the integer subscripts indicate the angular momentum
projection quantum number $m$ along the $z$-axis.
Because the standing laser field has only circular components,
and the Clebsch-Gordan coefficient between $m_e=0$ and $m_g=0$
states is zero, the atoms are rapidly pumped to the $\Lambda$
subsystem of the whole state structure;
thus, in this level configuration, the atoms are trapped to the ground 
substates which
have an angular momentum quantum number $m_g=-1$ and $m_g=+1$.
The excited state with $m_e=0$ provides a way for cooling
optical pumping cycles between the two trapping ground
substates.

After the rotating wave approximation the Hamiltonian for
the atomic system interacting
with the laser field given in Eq.~(\ref{eq:Efield})
is
\begin{equation}
       H =  \frac{p^{2}}{2M}
       - \hbar \delta P_{e}
        + {V} 
        + {U} \label{eq:HBlue}.
\end{equation}
Here, $P_{e} =\sum_{m=-1}^{1} |e_m \rangle~
\langle e_m|$ is the projection operator, and the interaction between a single atom and the
field is
\begin{eqnarray}
       V&=& i\frac{\hbar\Omega}{\sqrt{2}} \sin(kz)
       \left\{|e_{0} \rangle~\langle g_{-1}| +
       |e_{1} \rangle\langle g_{0}|\right\}
 \nonumber \\
       &&
 +\frac{\hbar\Omega}{\sqrt{2}}\cos(kz)
       \left\{|e_{-1} \rangle\langle g_{0}| +
       |e_{0} \rangle\langle g_{1}|\right\} +H.c.,
       \label{eq:VAtomLaserBlue}
\end{eqnarray}
where $z$ is the position operator of the atom, and
the Rabi frequency $\Omega$ is defined as
\begin{equation}
\Omega = 2 d {\cal E}_0 / \sqrt{2} \hbar.
\end{equation}

The interaction with a magnetic field in Eq.~(\ref{eq:HBlue})
is
\begin{equation}
     {U}=
     \sum_{i}m_i\hbar\Omega_{B_{i}} |i\rangle\langle i|,
\end{equation}
where the sum over $i$ includes all the ground and excited states, and
the Zeeman shift factors $\Omega_{B_{i}}$ for the
two trapping  ground substates $m=\pm1$ are equal.

The level structure which we have used in our studies for
a blue-detuned lattice can be found in $^{87}$Rb, which has $F=1$
hyperfine states  for both the 5S$_{1/2}$ ground state and
the 5P$_{1/2}$ excited state.
This is actually the element and the level scheme
which has been used in the blue-detuned lattice experiment of
Hemmerich {\it et al.} \cite{Hemmerich95a}, even though in their
case the lattice structure and the orientation of the magnetic
field differs from what is presented here.

The reason for the choice we have made for the used level structure
and an antiparamagnetic regime of the Grynberg-Courtois
lattice is in their simplicity. The cooling mechanism resembles
the traditional Sisyphus cooling, making it more relevant
to compare the results between the red and blue-detuning
studies. Moreover, it is necessary
to use only three levels of the $\Lambda$-subsystem
instead of all the six levels of a single atom.
Thus the number of product state basis vectors
for interaction studies between the atoms can be reduced
in the blue-detuned case. This reduces the 
computational resource requirements and 
speeds up the
simulations when compared to red-detuned case (which
has to use all the six substates of a single atom for
the interaction studies). Thus for blue-detuned
lattices it is easier to make a wider exploration
of parameter space, if required.

\subsection{Basic theoretical approaches}

Because of the polarization
gradients, the laser couples the 
multitude of Zeeman substates of the atom in a 
position dependent way and the spontaneous
emission caused by the coupling to the vacuum
plays a crucial
role in the optical pumping process.
Thus, to describe the atomic motion in optical lattices one
has to solve the problem of a multi-level atom
coupled to a monochromatic laser field and
to a quantized electromagnetic environment in its vacuum state.

It is possible to treat the external laser field classically
since the fields which are considered weak from
the lattice point of view still 
contain
a large number of 
photons. Typical laser intensities used in experiments are
a few mW/cm$^2$.  
The treatment of the interaction between a classical field and an atom
is typically done with the rotating wave approximation,
which neglects the terms that do not conserve the total energy.

The general form of the task is to solve the master equation
for the density matrix $\rho$ of the atomic system
\begin{equation}
i\hbar\frac{d\rho}{dt}=\left[ H,\rho \right] +{\cal L}_{rel}\left[ \rho \right],
\label{eq:Master}
\end{equation}
where $H$ is the system Hamiltonian, see 
Eqs.~(\ref{eq:Hred}), (\ref{eq:HBlue}), and ${\cal L}_{rel}$ includes the spontaneous
emission part due to the coupling to the environment.

It is extremely difficult to find the exact analytical solution for this
equation even in
the case of the simplest atomic level schemes
used for optical lattices. One can try to find
approximations which allow an analytical treatment,
or the combination of analytical and numerical 
calculations to Eq.~(\ref{eq:Master}). Another
possibility is to simulate the optical lattice
system on a computer, especially,
simulations by the Monte Carlo wave-function (MCWF) method
may provide a convenient way to obtain the solution
of Eq.~(\ref{eq:Master}) 
\cite{Dalibard92a,Molmer93a,Molmer96a,Plenio98a}.
In the case of laser cooling,
the combinations of analytical
and numerical treatments can treat 2D case
\cite{Berg-Sorensen93a},
but the 3D problem has been solved so far only by
the MCWF simulations \cite{Castin95a}.

A common feature for analytical treatments
of optical lattices
is the adiabatic elimination of the excited states, thus
reducing the number of the levels in the problem
at hand \cite{Guidoni99a}. 
One can then try to solve the master
equation directly by numerical integration, introduce
further approximations, or exploit the translational
symmetry properties of the system \cite{Castin91a}.

The MCWF method was originally developed for
the problems in quantum 
optics,
where
in many of the cases
the direct quantum-mechanical
solution of the system density matrix is
very difficult or impossible to 
obtain. The method has been applied e.g. to the resonance fluorescence
spectrum of 1D optical molasses \cite{Marte93a}
and  to 3D laser
cooling \cite{Castin95a}.
For more examples, see Ref.~\cite{Plenio98a}. 
By now the method
has also been applied outside the field
of quantum optics, 
e.g. into transport problems
in condensed matter physics~\cite{Badescu01a}.
The key idea of MCWF method is the generation
of a large number of single wave function realizations
which include stochastic quantum jumps of the system studied.
The final result for the system density matrix and the system properties
can then be calculated as ensemble averages of single realizations.
Section \ref{subsec:MCWF} presents the method in more detail.

\begin{figure}[tb]
\centering
\psfig{figure=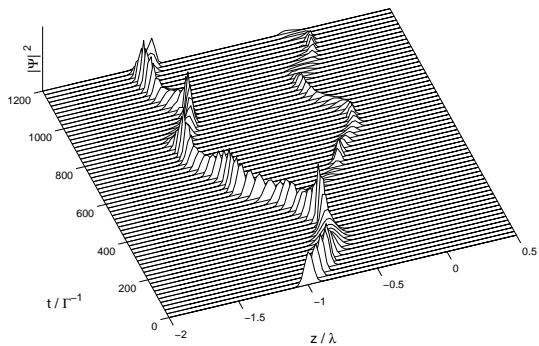,scale=0.4}
\caption[f3]{\label{fig:Branching}
Branching of a wave packet. 
The time evolution of the total atomic wave packet
in position
space is shown. In addition to treating spontaneous decay
in a rigorous way, the full quantum-mechanical MCWF method
can also account for the branching of the wave packet
in optical lattices.}
\end{figure}

We have chosen for our collision studies in optical lattices 
the MCWF method since it has been a widely used benchmark
method for various semiclassical theories for
cold collisions in MOTs \cite{Suominen98a}.
The method gives a full quantum-mechanical
description of the atomic system (the external
laser field is still described classically) and treats spontaneous
decay in a rigorous way. A semiclassical version
of the Monte Carlo (MC) method has also been developed for lattice
studies \cite{Petsas99a} and has been applied
e.g. to study anisotropic velocity distributions
in 3D dissipative optical lattices \cite{Jersblad03a}.
This variant describes the external
atomic degrees of freedom classically, and has limited applicability when
the spread of the wave packet influences the dynamics of the
system. This is the case when an atom is tightly localized
into a lattice site and the spread of the packet affects essentially the
optical pumping rate.
Moreover, the semiclassical approach can not treat, e.g., the
branching of the wave packet in optical lattice shown in 
Fig.~\ref{fig:Branching}. For the branching of the packets, see e.g.~discussion
in Ref.~\cite{Greenwood97a}.

\section{Cold collisions between laser cooled atoms
in the presence of near-resonant light}\label{sec:colcol}

The thermal velocities of the atoms in a laser cooled gas are 
on the order of centimeters per second, roughly four
orders of magnitude less than in a room temperature gas.
Consequently, the collision dynamics
is strikingly different within the two
temperature regimes.
The kinetic energies involved in the 
cold collision process are on the order or less than the energy of the
atomic linewidth.
For example, in the case of cesium the linewidth of a typical cooling 
transition is $2400$E$_r$ and
a typical collision energy is around $1600$E$_r$.
Slow atomic motion indicates that the decay time of the excitation
becomes  small compared to the time scale of the total collision
dynamics.
This allows new phenomena in cold collision processes
to affect the thermodynamics of the cold atomic cloud
\cite{Suominen96a,Weiner99a}.

A general categorization of the dynamical interactions
between atoms
can be made by considering collisions occurring between two ground
state or between one ground and one excited state atom.
Here we describe binary collisions between
atoms of same element,
i.e., a homonuclear diatomic molecule,
and the emphasis is on the collisions
between a ground and an excited state atom
which is the relevant scheme
for dissipative optical lattices.

In the presence of near-resonant light the excitation of a quasimolecule
formed by the colliding atoms may occur at an
extremely long range, even on the order of few thousands
of Bohr radii a$_0$.  Molecular potentials at these
long ranges are usually labeled by the Hund's case
(c) notation, where the component of the total electronic
angular momentum along the internuclear axis 
is a good quantum number.
The electron clouds of the colliding atoms do not overlap
at long range, 
and the dominant interaction between the atoms is the resonant dipole-dipole
interaction. We give a description of the resonant
dipole-dipole interaction in Section~\ref{subsec:DipDip}, 
earlier calculations for alkali-metals can be found in 
Refs.~\cite{Movre77a,Julienne91a}.

The long range 
properties of molecular states and the sign of the detuning of the 
laser with respect the appropriate electronic transition
define the possible consequences of collisions.
A repulsive or an attractive character of the molecular potential
at long range arises due to the relative orientation of the dipole moments
of the colliding atoms.
In the case of a red-detuned laser,
the resonance or Condon point $r_c$,
occurs for an attractive state. For a blue-detuned laser
$r_c$ occurs for a repulsive state.
Depending on the magnitude of the detuning and the strength of
the coupling
laser, off-resonant excitation to a non-resonant
state may also play a role \cite{Piilo02b}. 

A very active research field of its own is the
photoassociation of laser cooled atoms
to molecules. Photoassociation is typically done by using
a large red detuning of the laser 
so that free atom pair is excited at $r_c$
to a well defined bound molecular vibrational state.
Especially photoassociation in an atomic 
Bose-Einstein condensate has attracted wide interest
recently 
\cite{Drummond98a,Javanainen99a,Wynar00a}.
We will not discuss further photoassociation here
but the reader can find presentations
of the field e.g. from Refs.~\cite{Weiner99a,Kostrun00a}.

\begin{figure}[tb]
\centering
\psfig{figure=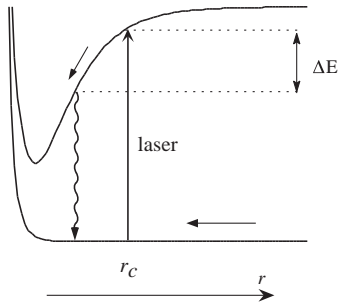,scale=0.4}
\caption[f4]{\label{fig:RadiativeHeating}
A schematic view of radiative heating of colliding atoms.
The quasimolecule is excited at 
the Condon
point $r_c$ and accelerated on the 
upper level before spontaneous 
decay terminates
the process. The consequent kinetic energy gain
is noted by $\Delta E$
 }
\end{figure}

\subsection{Radiative heating
by red-detuned light}\label{subsec:RH}

With red-detuned light, the population of the quasimolecule
formed by the two collision partners may get resonantly excited into
an attractive excited state at the Condon point $r_c$
(see Fig.~\ref{fig:RadiativeHeating}). The relative velocity
of the atoms increases due to the acceleration on the
attractive state until spontaneous emission
terminates the process. When the atoms have again bounced
apart due to the short range repulsion in the ground state,
the pair may lose some of the gained kinetic
energy in the reverse process, but to lesser degree. 
The overall effect is the heating 
of the colliding pair, and the escape of the atoms from the trap
if the total gain in kinetic energy is large enough \cite{Weiner99a}.

Figure~\ref{fig:RadiativeHeating} shows a semiclassical
(SC)
schematic view of the process. In some of the
parameter regimes
SC descriptions, such as the Landau-Zener level crossing
model, can be used \cite{Suominen98a}.
When the SC models fail, full quantum-mechanical
methods are needed. For example, MCWF simulations
\cite{Holland94a,Holland94b}
can be used as a benchmark method for
simpler analytical semiclassical calculations.
It should be emphasized that it is difficult in SC
models to account for population
recycling, which means that once-decayed population may get
re-excited in strong laser fields. 
A comparison between various methods
and their application range is given in Ref.~\cite{Suominen98a}.

\begin{figure}[tb]
\centering
\psfig{figure=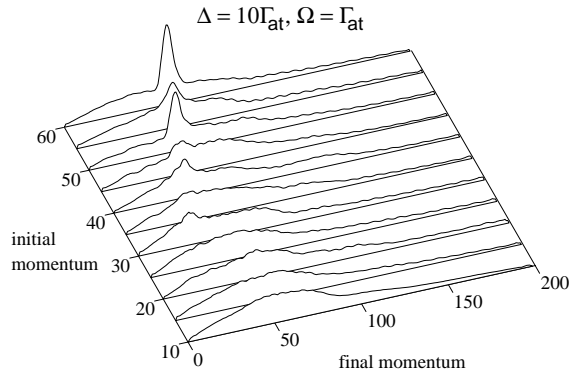,scale=0.4}
\caption[f4]{\label{fig:RhInMot}
An example of a radiative heating
study in MOT from Ref.~\cite{Holland94b}. 
The final momentum
distributions for various initially 
Gaussian momentum distributions
are shown. The initial momentum refers to the $\langle p\rangle$
of the narrow initial momentum distributions.
The effects of radiative heating,
the spreading of the distributions,
and the momentum increase is clearly visible,
especially
for low initial relative momentum. For higher initial 
momentum
some character of the initial distribution
is still preserved in the post-collision distribution.
 }
\end{figure}

Most of the radiative heating studies done so far
have used a simple two-state description with 
one ground and one excited molecular electronic state. 
An example of MCWF simulation results for MOT
from Ref.~\cite{Holland94b} is
shown in Fig.~\ref{fig:RhInMot}. The results
demonstrate radiative heating via the spreading of the momentum distributions.
The effect clearly becomes stronger as the initial collision momentum, i.e., cloud temperature, decreases.
We emphasize that our lattice studies include
many
attractive and repulsive states simultaneously, see Section
\ref{subsec:DipDip}. 

In addition to radiative heating, atoms may also
escape from the trap by the fine-structure
change mechanism.
If the population survives on the excited state for small enough relative distance
between the two atoms, the point where
two fine-structure  states have a crossing may be reached.
Now, if the pair comes out from the 
collision in an energetically 
lower fine-structure state than the one in which
they entered the collision,
the pair gains kinetic energy by the amount corresponding
to the fine-structure splitting of 
the atomic electronic states
at large $r$. This energy difference is usually
large compared to the trap depths, and consequently
in this case the atoms escape from the trap \cite{Suominen96a,Weiner99a}.

We deal in our studies with very small detunings, 
a few atomic linewidths 
only, and strong laser fields.
It is therefore reasonable to assume that
the effect of fine-structure changing collisions
on the cloud temperature 
is very small compared to the effect of radiative heating 
processes, and we neglect the fine-structure change loss mechanism
in our lattice studies. The assumption is also supported by the fact that
there can be an order of magnitude difference
between the fine-structure state crossing point and $r_c$ for the
small detunings we have used.

\subsection{Optical shielding by blue-detuned light}\label{subsec:Shielding}

\begin{figure}[t!]
\centering
\psfig{figure=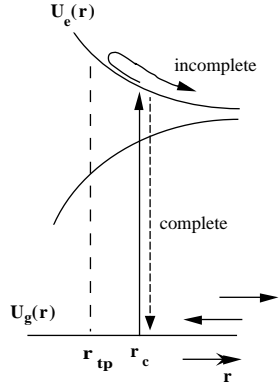,scale=0.5}
\caption[f3]{\label{fig:Shielding}
A schematic semiclassical presentation of optical shielding. The
quasimolecule is excited resonantly to the repulsive molecular state
$U_e$ at the
Condon point $r_{c}$. Then it reaches the classical turning point $r_{tp}$, and
is finally transferred back to the ground state $U_g$ when arriving at $r_{c}$ again.
If the transfer back to the ground state is not complete, the atom pair may
gain kinetic energy as it is further accelerated by the excited state
potential. In this case shielding is incomplete and the collision is
inelastic. If the population transfer between the states is adiabatic,
shielding is complete and the collision between the atoms is elastic. }
\end{figure}

When blue-detuned light is used, 
the resonant excitation at the Condon point occurs to a repulsive
excited quasimolecular state. This makes 
it possible to shield the atoms from close encounters \cite{Weiner99a},
see Fig.~\ref{fig:Shielding}.
If the shielding is efficient, collisions between atoms may become
completely  elastic.
The mechanism would obviously
be useful for preventing loss
of atoms in optical lattices formed with a near-resonant blue-detuned light.

In an optical shielding process, the resonantly
excited quasimolecule population reaches 
the classical turning
point on the repulsive excited state and the atoms begin
to move apart again. The shielding
becomes complete if all the population has been excited,
no spontaneous decay has occurred, and all the 
population returns resonantly to the ground state at
the Condon
point. In this case, collisions
become elastic
(when photon recoil effects are ignored),
and no heating or escape occurs due
to the inelastic processes. Moreover,
the ground state is emptied at
a relatively long range, and 
no population reaches short distances where
unwanted processes are possible,
such as hyperfine state changing collisions. Thus, 
the possibility to use optical shielding in an efficient
way allows
to increase the occupation density
of the dissipative lattice, in addition to the benefit of
reducing the rate of scattered and reabsorbed photons
and the consequent effects.

\begin{figure}[tb]
\centering
\psfig{figure=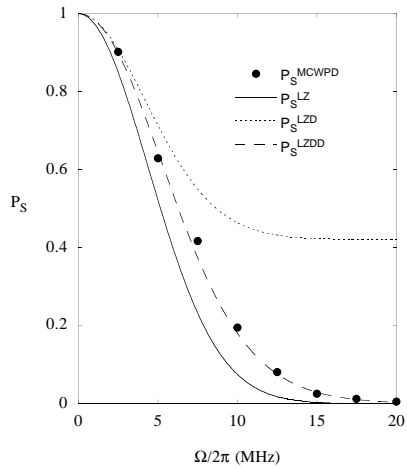,scale=0.4}
\caption[f4]{\label{fig:OsInMot}
An example of an optical shielding
study in a MOT from Ref.~\cite{Suominen95a}. 
Shown is the shielding measure $P_S$
as the function of the Rabi frequency $\Omega$.
$P_S$ describes the ground state population
flux at very short range. The dots are results
of the Monte Carlo simulations and the lines
presents the results of various 
semiclassical Landau-Zener approaches (see text).
When the laser field is strong enough,
the ground state is effectively emptied
before the atoms have a chance to approach
close to each other. This  is demonstrated  by the decreasing
value of $P_S$ for increasing Rabi frequencies. 
 }
\end{figure}

In the past the MCWF simulations have 
described the efficiency of the shielding process
in a MOT 
by a shielding measure $P_S$, which essentially
describes the flux of the ground state population to 
the short range beyond the Condon 
point $r_c$ 
\cite{Suominen95a,Suominen96c,Suominen96b},
see Fig.~\ref{fig:OsInMot}. In the case of optical lattices,
a more descriptive result
is the momentum distribution in a steady state
compared between interacting and non-interacting
atoms \cite{Piilo02b}. 

In addition of the MCWF simulations, Fig.~\ref{fig:OsInMot} displays
also various semiclassical Landau-Zener
results for the shielding measure. In the basic Landau-Zener approach (LZ),
 the calculation
of the excitation probability is based on the linear level crossing model 
and assuming narrow initial wave packet in momentum
space \cite{Suominen95a,Landau32a,Zener32a}. 
The Landau-Zener model with decay (LZD) takes into account the exponential
decay of the population from the excited electronic state 
\cite{Suominen95a} whereas
the Landau-Zener model with delayed decay (LZDD) accounts also for the
re-excitation of the once decayed population due to the strong laser
field \cite{Suominen95a}.

In this Section we have described radiative heating
and optical shielding processes by using the molecular
state description. This is how the treatment has usually
been done in the past when studying the heating and loss of atoms in
MOTs. In our calculations and simulations we use the two-atom product
state  presentation instead. Our aim is to study the effect
of collisions in a near-resonant optical lattice. Most of the photon scattering
still occurs when the atoms are outside the range
of a binary interaction. It would be an unnecessary complication
to describe photon
scattering and quantum jumps in a lattice
by using a molecular basis. 
Moreover, we give a simultaneous description
for laser cooling and collision processes. Hence, in our case
the molecular basis is only used for a
qualitative description
of radiative heating and optical shielding processes,
but it does not appear directly in our calculations.

\section{Collisions in optical lattices}\label{CollisionsInLattice}

In the past cold binary collisions between atoms have been widely studied
under conditions which correspond to the atoms trapped in magneto-optical
traps \cite{Suominen96a,Weiner99a,Burnett95a}.
Previous cold collision research has concentrated on
the effects of inelastic collisions
on the properties of an atomic cloud,
but has neglected
the co-existence of the cooling processes.

We are dealing with near-resonant optical lattices, and
we have combined in a single framework the cooling
and cold collision 
dynamics.
Thus our approach includes simultaneous dynamical 
processes of cooling, trapping,
and collisions in atom gas.

For far off-resonant lattices the
possibility to control the cold collisions coherently
has been proposed \cite{Jaksch99a}
and realized in an experiment \cite{Mandel03a}.
This allows the
creation of an entanglement between the atoms,
a step towards quantum computing
in optical lattices \cite{Jaksch04a}. Moreover,
the observation of 
superfluid--Mott-insulator phase
transition helps in filling in
the sites of 
the far-off resonant lattice
by controlled particle number \cite{Greiner02a}. 
For dissipative optical
lattices, the creation of a double 
optical lattice opens interesting
prospect for cold collision studies 
in the presence of near-resonant light \cite{Ellmann03a}.

Radiative heating and optical shielding studies in MOTs
typically use the molecular state description of the binary atomic system.
After choosing the specific excited molecular state
and doing the partial wave expansion, usually only the lowest
relative angular momentum ground and excited state
are accounted. The other option has been to consider
independent pairs
of partial wave states, neglecting the coupling between the
pairs in a weak field approximation \cite{Holland94b}.
Thus the descriptions in the past have been two-state models, neglecting
the multitude of internal states of the atoms,
and
without the position dependent coupling between the
atoms and the laser field. For optical lattices the internal
states of the atoms have to be accounted because of the spatial periodicity
of the coupling caused by the polarization gradients
of the laser field. 

We use the two-atom product state basis \cite{Cohen-Tannoudji77a}.
For two six-level
atoms in the red-detuned case
this means $36$ basis states. If the basis is transformed
into a molecular one, there are manifolds of attractive and
repulsive molecular states buried in our description. We are
forced to
do the calculations in one dimension since
the limited amount of computational 
resources. It is also worth noting that the simple
atomic level schemes we consider, do not allow
to account for the spatially dependent adiabatic
couplings between the atomic states which play
a role in the system dynamics for more complex
level schemes~\cite{Grynberg01a}.

A resonant dipole-dipole interaction between the atoms 
in optical lattices
has been studied with nondynamic approaches
\cite{Goldstein96a,Boisseau96a,Guzman98a,Menotti99a,Menotti99b}.
Usually
these studies assume fixed positions for the atoms
and concentrate on the mean-field type descriptions
of the lattice system. These approaches neglect the dynamical
nature of the cold collisions, and the
inelastic processes of radiative heating
or incomplete optical shielding. Our approach includes 
the dynamical processes of cooling and
cold collisions in the same 
framework. 
It is important to note that once
the atoms are localized into the optical lattice sites, they are 
still able to 
move around in the lattice. For shallow lattices this
is because the quantum-mechanical 
tunneling probability between the lattice sites is not negligible. For
deep optical lattices, which are more relevant to
our case, the atomic motion between the wells may
be induced by the recoil effects combined
with the optical pumping process. 

For the lattice parameters we use, we have noticed
that the inter-well effects are negligible and
our interest lies in the case when two atoms
end up in the same lattice site and collide. The intra-well
collision
partners may then gain kinetic energy due 
to an inelastic collision and escape from the lattice. In the
case of optical shielding, the possibility of making the 
collisions elastic and preventing the atoms from close
encounters would also  prevent the atoms to escape from  the
lattice.

Numerical simulations are extremely heavy, especially
in the case of red-detuned lattices where the level scheme
can not be simplified. For the required computer resources,
see Appendix A. We are forced to make some
simplifications to our model, for details see \cite{Piilo02a}.
Most importantly, 
we have to neglect the reabsorption of the scattered
photons. With increasing atomic density, reabsorption
may heat the atomic cloud and cause radiation pressure to outward
direction from the trap center \cite{Metcalf99a}, limiting the
achievable atomic densities.
In the blue-detuned lattice, the number of scattered photons
is largely reduced because the center
of each lattice site corresponds to a completely dark point
in space. For red-detuned lattices we
merely describe the effects of collisions on the thermodynamical
properties of the cloud. The full thermodynamics
is not described since we neglect reabsorption.
This poses some limitations for the applicability of our
results in red-detuned lattices, but for the blue-detuned
case our description is close to the complete 
thermodynamical description because the
scattering is generally low.

In the following Subsection we describe how we calculate
the interaction matrix elements between two
multistate atoms, especially the resonant 
dipole-dipole interaction matrix elements. 
The whole
problem is then formulated by using the MCWF method.
For details, see Ref.~\cite{Piilo02a}.
The Section ends
with the presentation of the central results.

\subsection{Resonant dipole-dipole interaction}\label{subsec:DipDip}

One of the early treatments for resonant dipole-dipole
forces between two atoms was given already at the end
of the 30's \cite{King39a}, and the retardation effects were
discussed almost a
decade later \cite{Casimir48a}.
Later on, Lenz and Meystre considered the
resonant dipole-dipole interaction (DDI) for two two-level 
atoms in a standing-wave field \cite{Lenz93a}.
Our derivation follows their approach and makes 
the generalization
to the multilevel atom case.

The DDI is the first interaction to come
into play when the colliding
atoms approach each other in dissipative
optical lattice. 
Since the interaction between the atoms is mediated by the quantized
environment, the natural starting point is the
two-atom system master equation and
its damping part describing the coupling of
the system to the electromagnetic
environment \cite{Lenz93a}
[see also Eq.~(\ref{eq:Master})] 
\begin{eqnarray}
\dot{\rho}
&=& -\frac{1}{\hbar^2}
\int_{0}^{t} d \tau Tr_{f} 
\left\{
H_{sf}(t)H_{sf}(\tau)\rho_{sf}(\tau)
-H_{sf}(t)\rho_{sf}(\tau)H_{sf}(\tau)
 \right.
\nonumber \\
&&
\left.
 -H_{sf}(\tau)\rho_{sf}(\tau)H_{sf}(t)
+\rho_{sf}(\tau)H_{sf}(\tau)H_{sf}(t) 
\right\},
\end{eqnarray}
where $\rho$ is the reduced density matrix of the two-atom system,
$\rho_{sf}$ the density matrix of the two-atom system and the field, 
$H_{sf}$ denotes the system-field interaction
Hamiltonian, 
and $Tr_f$ the trace over the field.

We expand the electromagnetic field in the standard
way 
\begin{eqnarray}
{\bf E}({\bf r}_\alpha)&=&{\bf E}^+({\bf r}_\alpha)+
                                    {\bf E}^-({\bf r}_\alpha) \nonumber \\
{\bf E}^+({\bf r}_\alpha)&=&\sum_{\bf {k}} i{\cal E}({\bf k}) {\bf a}_{\bf k}
e^{i{\bf k\cdot{\bf r}_\alpha}} \nonumber \\
{\bf E}^-({\bf r}_\alpha)&=& 
\left( {\bf E}^+({\bf r}_\alpha) \right) ^{\dag},
\end{eqnarray}
where ${\bf a_{\bf k}}$ is the annihilation operator
for mode ${\bf k}$,
${\bf r}_\alpha$ denotes the position
of atom $\alpha$, and 
\begin{equation}
{\cal E}({\bf k}) = \sqrt{\frac{2 \pi \hbar \omega({\bf k})}{V}}{\bf \epsilon}_j({\bf k}),
\end{equation}
where ${\bf \epsilon}_j$ is the polarization vector
and $V$ the quantization volume.

We use  the center of mass and relative
coordinates of the atom pair
\begin{eqnarray}
{\bf R} &=& \frac{{\bf r}_1+{\bf r}_2}{2} , ~~{\bf r}  = {\bf r}_2-{\bf r}_1,
\end{eqnarray}
and the notation
\begin{eqnarray}
     S_{+,q}&=&\sum_{m}CG_{m}^{q}
      \left( |e_{m+q}\rangle_1~_1\langle g_m| + 
     |e_{m+q}\rangle_2~_2\langle g_m| \right)
\nonumber \\
\Delta S_{+,q}&=&\sum_{m} CG_{m}^{q}
      \left( |e_{m+q}\rangle_1~_1\langle g_m| - 
     |e_{m+q}\rangle_2~_2\langle g_m| \right),
\end{eqnarray}
where $CG_{m}^{q}$ are the appropriate Clebsch-Gordan coefficients,
$q$ is the
polarization label in the spherical basis, and sub-indices label the two
atoms.
The interaction between the two-atom system and the
vacuum electromagnetic field can now be written as
\begin{eqnarray}
H_{sf}&=&i\sum_{{\bf k},{\bf \epsilon}_j}
\sqrt{\frac{2 \pi \hbar \omega({\bf k})}{V}} 
d\left\{ \epsilon_{j,+}\left[\cos\left({\bf k}\cdot\frac{\bf r}{2}\right)S_{+,+}
-i \sin\left({\bf k}\cdot\frac{\bf r}{2}\right)\Delta S_{+,+} \right]  \right. \nonumber \\
&& +\epsilon_{j,0}\left[\cos\left({\bf k}\cdot\frac{\bf r}{2}\right)S_{+,0}
-i \sin\left({\bf k}\cdot\frac{\bf r}{2}\right)\Delta S_{+,0} \right]  \nonumber \\ 
&&
\left.
+\epsilon_{j,-}\left[\cos\left({\bf k}\cdot\frac{\bf r}{2}\right)S_{+,-}
-i \sin\left({\bf k}\cdot\frac{\bf r}{2}\right)\Delta S_{+,-} \right]   \right\} 
 e^{i{\bf k}\cdot{\bf R}}e^{i\omega_0t} {\bf a}_{{\bf k},j} 
\nonumber \\
&&
+H.c.,
\end{eqnarray}
where $\epsilon_{j,q}$ is the projection 
$\epsilon_{j,q}={\bf \epsilon}_j\cdot{\bf \epsilon}_q$
on the spherical basis ${\bf \epsilon_0},
{\bf \epsilon}_{\pm}$, and $d$ the 
dipole moment of the atomic transition.

One can identify the DDI interaction terms between
the atoms as those
having $\langle n_\omega+1\rangle=1$ where $n_\omega$ is the number
of photons in the mode of the environment with 
mode frequency $\omega$.
In another words, the average photon number 
in the interaction process is zero
since the DDI interaction can be viewed as an exchange
of excitation between the two atoms via
the environment vacuum field. 
After lengthy analytical calculations 
following \cite{Lenz93a},
and using the arguments
from Ref.~\cite{Berman97a},
one can write down the expression for the three-dimensional
resonant dipole-dipole interaction as

\begin{eqnarray}
     V_{dip} &=&- \frac{3}{8} \hbar \Gamma \left\{
     \frac{1}{3} \frac{\cos q_0 r}{q_0 r}
     \left[1-2P_2 (\cos \theta_r)\right]
 \left({\cal S}_{++}{\cal S}_{-+} +
     {\cal S}_{+-} {\cal S}_{--}  - 2 {\cal S}_{+0}{\cal S}_{-0} \right)
 \right.    
\nonumber \\
     && - 2\left[\frac{\sin q_0 r}{(q_0 r)^2} + \frac{\cos q_0 r}{(q_0 r)^3}
     \right] P_2(\cos \theta_r)  
\left( {\cal S}_{++}{\cal S}_{-+} +
     {\cal S}_{+-} {\cal S}_{--}  - 2 {\cal S}_{+0}{\cal S}_{-0} \right)
     \nonumber \\
     && + \frac{1}{3} \left[-\frac{\cos q_0 r}{q_0 r}
     +3\left(\frac{\sin q_0 r}{(q_0 r)^2} + \frac{\cos q_0 r}{(q_0
r)^3}\right)\right] \times
     \nonumber \\
     && \left[ \frac{1}{\sqrt{2}} P_{2}^{1} (\cos \theta_r) \cos \phi_r
 \left(-{\cal S}_{++} {\cal S}_{-0} + {\cal
S}_{+0}{\cal S}_{--}
     -{\cal S}_{+0}{\cal S}_{-+}+{\cal S}_{+-}{\cal S}_{-0}\right)
 \nonumber 
\right.
\\
     && +  P_{2}^{2} (\cos \theta_r) \cos 2\phi_r 
 \left. \left.
 \left({\cal S}_{++} {\cal S}_{--}+{\cal
S}_{+-}{\cal S}_{-+}
     \right)\frac{}\! \right] \right\},
\end{eqnarray}
where $q_0=\omega_0/c$, $P_{2}$ is Legendre polynomial, $P_{m}^{n}$
are the associated
Legendre functions, and $\Gamma$ is the linewidth of the atomic excited state. 
The angles $\theta_r$ and $\phi_r$ are the angles of the
relative coordinate ${\bf r}$ in the spherical basis. We have also
introduced the
operators
\begin{equation}
      {\cal S}_{+q} {\cal S}_{-q'} \equiv
      \left( S_{+,q}^{1}S_{-,q'}^{2} + S_{+,q}^{2} S_{-,q'}^{1}
      \right), \label{eq:S+qS-q}
\end{equation}
where $S_{-,q}^{\alpha}=\left( S_{+,q}^{\alpha}\right)^{\dagger}$
and
\begin{equation}
      S_{+,q}^{\alpha}=\sum_{m=-J_g}^{m=J_g} CG_{m}^{q}
      |e_{m+q}\rangle_\alpha ~_\alpha\langle g_m|.
      \label{eq:S+q-alpha}
\end{equation}
Here  $\alpha$ labels one of the two
atoms.

\begin{figure}[tb]
\centering
\psfig{figure=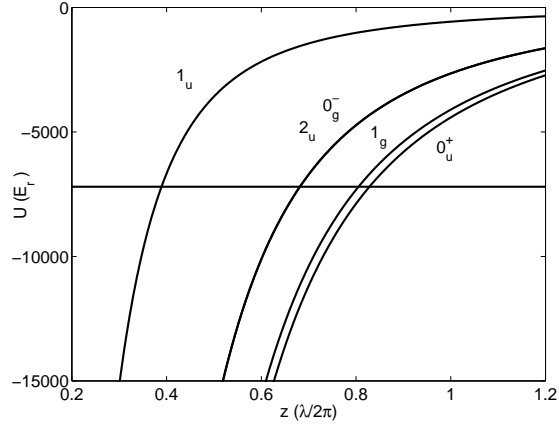,scale=0.4}
\caption[f5]{\label{fig:MolPots}
The energy shifted ground state and the attractive excited state [labeled by
the Hund's case (c) notation] molecular potentials of Cs$_2$ 
for $\delta=-3.0 \Gamma$.
The repulsive potential manifold is not shown.
}
\end{figure}

If the two atoms are positioned on the $z$-axis, the DDI potential
reduces to the
one-dimensional potential
\begin{eqnarray}
      V_{dip}^{axis}&=&\frac{3}{8}\hbar\Gamma \left\{ \frac{1}{3}
      \frac{\cos q_0 r}{q_0 r}
      +2\left[ \frac{\sin q_0 r}{(q_0 r)^2} + \frac{\cos q_0 r}{(q_0 r)^3}
      \right] \right\} 
\times \nonumber \\
      &&~~~
\left( {\cal S}_{++}{\cal S}_{-+} + {\cal S}_{+-}{\cal S}_{--}
      - 2{\cal S}_{+0}{\cal S}_{-0} \right).
      \label{eq:VDipAxis}
\end{eqnarray}
It is worth noting that the interaction potential (\ref{eq:VDipAxis}) includes
the retardation effects.
By diagonalizing $V_{dip}$,
it is possible to obtain the molecular potentials shown
in Fig.~\ref{fig:MolPots}. We use 
the molecular basis
occasionally for the qualitative description
of the collision processes and 
emphasize that the
calculations are done in the two-atom product state basis.

\subsection{Monte Carlo wave-function formulation}\label{subsec:MCWF}

We use a variant of the Monte Carlo (MC) method which was
developed by Dalibard, Castin, and
M{\o}lmer \cite{Dalibard92a,Molmer93a,Molmer96a}.
The core idea of the Monte Carlo wave-function (MCWF) 
method is the generation of a large
number of single wave
function realizations including stochastic quantum jumps of the system studied.
Quantum jumps occur to the available decay channels of
the system whose environment is continuously monitored. In our case,
detection of a photon 
corresponds to a quantum jump
from an internal excited electronic 
state to the ground electronic state of an atom in an optical lattice.
Solutions for the steady state density matrix and system properties can be
calculated as ensemble averages of single wave-function realizations. 

In general, we look for the solution
of the master equation (\ref{eq:Master})
whose relaxation part is in the so called
Lindblad form \cite{Lindblad76a}
\begin{equation}
{\cal L}_{rel} \left[ \rho\right] = 
-\frac{1}{2}\sum_i \left( C_i^{\dag} C_i \rho + 
\rho  C_i^{\dag} C_i \right) +
\sum_i  C_i \rho C_i^{\dag}, 
\label{eq:rel}
\end{equation}
where the summation over $i$
include the system operators $C_i$
and their adjoints $C_i^{\dag}$ (we define these operators in a moment).

To unravel the appropriate master equation
by generating the realizations for the  MC ensemble,
one solves the time dependent Schr{\"o}dinger equation
\begin{equation}
     i \hbar \frac{\partial|\psi\rangle}{\partial t}=
     H|\psi\rangle. \label{eq:Schrodinger}
\end{equation}
In our case $|\psi\rangle$ describes the time-dependent
two atom wave-function in position space
\begin{equation}
     |\psi(z_2,t)\rangle  = \sum_{j_{1},j_{2},m_{1},m_{2}}
     \psi^{j_{1},m_{1}}_{j_{2},m_{2}}(z_2,t) |j_{1} {m_{1}}\rangle_{1}
     |j_{2} {m_{2}}\rangle_{2}, \label{eq:Psi}
\end{equation}
where $j_1$ and $j_2$ denote the ground and excited states of
atom $1$ and $2$ respectively, $m_1$ and $m_2$ the
$z$-component of the angular momentum. 
Due to the limited availability of computer resources 
we have to fix the position of atom 1 by setting $z_1=0$, and $z_2$ is then
the both position of the moving atom 2 in the lattice, as well as the relative coordinate \cite{Piilo02a}. 

The non-Hermitian Hamiltonian $H$ in Eq.~(\ref{eq:Schrodinger})
is
\begin{equation}
     H=H_{S}+H_{DEC} \label{eq:H},
\end{equation}
where the system Hamiltonian $H_S$ in our case 
includes the atom-laser interaction Hamiltonians
expanded in the two-atom Hilbert space, and the resonant dipole-dipole
interaction between the atoms, Eqs.~(\ref{eq:Hred},\ref{eq:HBlue},\ref{eq:VDipAxis}).

The non-Hermitian part includes the sum over the various allowed decay
channels $j$,
\begin{equation}
     H_{DEC}=-\frac{i\hbar}{2}\sum_{j}C_j^{\dagger}C_j, \label{eq:HDec}
\end{equation}
where $C_j$ are the jump operators corresponding to particular
decay channels and can be found out from 
the relaxation part of the master equation (\ref{eq:rel}). 

During a discrete time evolution step
of length $\delta t$ the norm of the wave function
may shrink due
to $H_{DEC}$. The amount of shrinking gives the probability of a
quantum jump to
occur during the short interval $\delta t$. Based on a random number one then
decides whether a quantum jump occurred or not. Before the next time
step is taken,
the  wave function of the system is renormalized. If and when a
jump occurs,
one performs a rearrangement of the wave function components
according to the jump
operator $C_j$, corresponding to the decay channel $j$, before renormalization of
$|\psi\rangle$.

For example,  
if we denote the jump of atom 1 from
$|e_{-1/2}\rangle_1$ to $|g_{-1/2}\rangle_1$ as channel 2
in our red-detuned lattice studies,
the jump operator in the product state basis for this jump is 
\begin{eqnarray}
     C_2&=&\sqrt{2/3}\sqrt{\Gamma}\left\{ |g_{-1/2} \rangle_1~|g_{-1/2}
     \rangle_2~_1\langle e_{-1/2}|~_2\langle g_{-1/2}|
 \right. \nonumber \\
     &&
+|g_{-1/2}\rangle_1~|g_{+1/2}\rangle_2~_1\langle
     e_{-1/2}|~_2\langle g_{+1/2}| \nonumber 
\\
     &&
+|g_{-1/2}\rangle_1~|e_{-3/2}\rangle_2~_1\langle
     e_{-1/2}|~_2\langle e_{-3/2}|
 \nonumber \\
     &&
+|g_{-1/2}\rangle_1~|e_{-1/2}\rangle_2~_1\langle
     e_{-1/2}|~_2\langle e_{-1/2}|  
\nonumber \\
&& 
+|g_{-1/2}\rangle_1~|e_{+1/2}\rangle_2~_1\langle
     e_{-1/2}|~_2\langle e_{+1/2}|  
\nonumber \\
     && 
\left.
\! +|g_{-1/2}\rangle_1~|e_{+3/2}\rangle_2~_1\langle
     e_{-1/2}|~_2\langle e_{+3/2}|\right\} \label{eq:C2}.
\end{eqnarray}
Here, the factor $\sqrt{2/3}$ is the Clebsch-Gordan coefficient
of the corresponding transition.
After applying
the jump operator $C_j$,  the wave function is still in a
superposition state, but
it has collapsed into a subspace of the product state basis
vectors, leaving only one
ground state
level component of the atom populated. 

In general, the jump probability into the decay
channel $j$ for each of the time-evolution step $\delta t$ is 
\begin{equation}
     P_j=\delta t \langle\psi|C_j^{\dagger}C_j|\psi\rangle. \label{eq:Jp}
\end{equation}
Thus, the jump probability 
for an example channel $2$ in Eq.~(\ref{eq:C2})
for each time step  
is
\begin{eqnarray}
     P_2&=&\frac{2}{3}\delta t \Gamma
     \left\{ 
|\psi^{e_{-3/2}}_{g_{-1/2}}|^2+
     |\psi^{e_{-3/2}}_{g_{+1/2}}|^2+
     |\psi^{e_{-3/2}}_{e_{-3/2}}|^2  \right. \nonumber \\
      && +\left.  |\psi^{e_{-3/2}}_{e_{-1/2}}|^2+
     |\psi^{e_{-3/2}}_{e_{+1/2}}|^2+
     |\psi^{e_{-3/2}}_{e_{+3/2}}|^2\right\} \label{eq:P1}.
\end{eqnarray}

Reference \cite{Piilo02a} presents in detail the implementation of
the MCWF method in our lattice studies. 
There is a large number of numerical problems
one has to solve for the MC implementation of two atoms in
a lattice, for a list and solutions see Ref.~\cite{Piilo02a}.
For general numerical tools, e.g. split Fourier or Crank-Nicholson
methods, to solve
time dependent Schr{\"o}dinger equation, see Ref.~\cite{Garraway95a}.

\subsection{Red-detuned lattices}
In this Subsection we present the main results from
Refs.~\cite{Piilo01a,Piilo02a} which deal with collision
dynamics in red-detuned lattices. The
main collision process in this case is
radiative heating, see Section \ref{subsec:RH}.

Once the atoms
are localized into the lattice sites, they are still able to move
around in the lattice. When the occupation density of the lattice
increases
one can ask what is the effect of collisions for the
cooling dynamics in optical lattices, and how the cold
collisions affect the atomic cloud once the atoms are localized. 

At the beginning of the efficient cooling period a large fraction
of atoms have higher kinetic energy
than the optical lattice modulation depth. 
Atoms then have a high mobility
and change their internal state frequently via the optical 
pumping cycles which cool them.
{\it A priori} one might assume that the possible
consequence of collisions would be a slowing of the cooling
process, heating, and escape of the atoms from
the lattice. Radiative heating
studies 
at low temperatures 
in MOTs show a smooth widening of the momentum
probability distribution corresponding to heating for 
large range of parameters \cite{Holland94a,Holland94b},
see Fig.~\ref{fig:RhInMot}. 
A similar effect might be expected to
occur in an optical lattice as well.

\begin{figure}[tb]
\centering
\psfig{figure=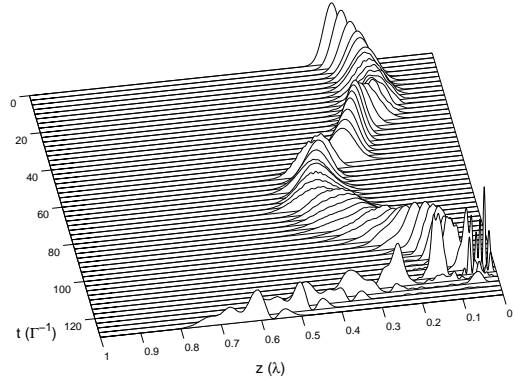,scale=0.4}
\caption[f12]{\label{fig:ex21}
An example of a single wave packet realization from
the full MC ensemble in position space
(note the direction of $z$-axis because
of the viewing angle).
At the beginning of  the time evolution 
the atom oscillates in the lattice well which has the center at
$z=0.25\lambda$. It then approaches point 
$z=0$ (the position of the fixed atom), collides with another atom, and
gains enough kinetic energy to be ejected from
the lattice.
}
\end{figure}

It turns out that the internal structure of the atoms 
and the spatial dependence of the atom-field coupling
changes the consequences of the collisions to some
extent. Lattice structure introduces selectivity into the collision
processes and atomic dynamics. In a lattice, the mobility of an atom
between the lattice sites depends essentially  on the kinetic energy,
especially once the atoms are localized
into the lattice sites. An atom,
which has a large oscillation amplitude
(corresponding to a large kinetic energy)
in the lattice site,
has a higher probability
to change its internal ground state by optical
pumping than an atom which is tightly localized
into the vicinity of the center of the potential
well. Rich dynamical
features of the wave packet, like breathing and oscillations in a single lattice site,
arise due to the fact that
the packet here describes a superposition
of the populations in the vibrational states of a lattice potential
\cite{Garraway95a}.

\begin{figure}[tb]
\centering
\psfig{figure=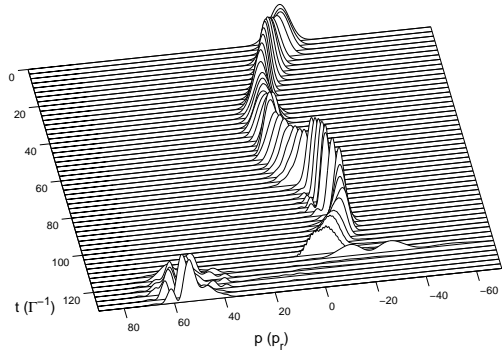,scale=0.4}
\caption[f12]{\label{fig:ex21k}
The same MC realization as in Fig.~\ref{fig:ex21}
but shown here in momentum space. 
The population is transferred
to the high
values of momentum due to the collision,
and the atom consequently escapes from the lattice.
}
\end{figure}

Since the high kinetic energy atoms are more mobile
compared to their low kinetic energy partners, the high
kinetic energy atoms have also a higher collision probability.
During the  same period of 
time the high-energy atoms change their internal ground
state and the corresponding optical potential more often
than tightly localized atoms.
Thus, the coverage of various lattice sites,
and the corresponding
collision probability, is higher for more mobile atoms. 

With these assumptions the consequence of
collisions might be simple heating
of the atomic cloud, or escape of the atoms from the lattice.
The essential ingredient for a large kinetic energy increase
is a high excitation probability of a quasimolecule. 
This depends on the curvature of the molecular
potentials at a resonant Condon point and especially
on the relative velocity between the colliding atoms.
The optical lattice modulation depth defines the initial velocity
distribution of the atoms when they start to move between
the lattice sites after localization. It turns out that in a 
lattice, detuned a few atomic linewidths below the
atomic resonance, and for lattice depths of a few hundred
recoil energies, the surroundings are very favorable
for strong excitation of the quasimolecule and the corresponding
large kinetic energy changing collisions. The consequence
is that the atoms mainly leave the lattice when colliding,
and the total effect is the ejection of the hot atoms
from the lattice. The ones which remain in a lattice 
have lower average kinetic energy per atom 
than in the low occupation density case of the
lattice when there is no need to account for
the interactions between the atoms.
Figures \ref{fig:ex21} and \ref{fig:ex21k}
show in position and momentum spaces, respectively, 
an example of a collision which
ejects the atoms from the lattice.

\begin{figure}[tb]
\centering
\psfig{figure=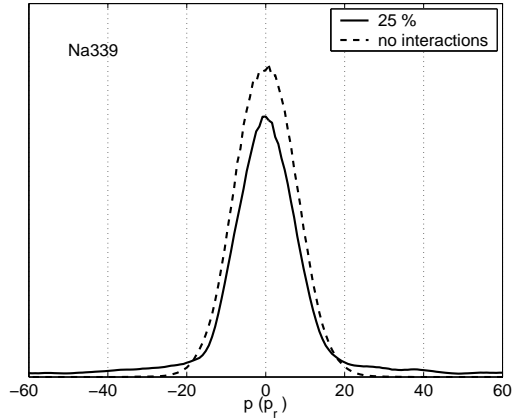,scale=0.4}
\caption[f12]{\label{fig:Narrowing1}
An example of momentum probability
distributions for interacting and non--interacting 
atoms in a red-detuned optical lattice \cite{Piilo02a}.
The momentum $p$
is expressed in the recoil unit $p_{r}=\hbar k_{r}$.
For interacting atoms the distribution
gets narrower compared to non-interacting atoms.
The results correspond to a sodium lattice with
lattice the depth $U_0=339E_r$, detuning $\delta=-3.0\Gamma$,
and Rabi frequency $\Omega=2.8\Gamma$.}
\end{figure}

The selective escape mechanism resembles evaporative cooling
used to produce BEC in magnetic traps. Here
the rethermalization of the remaining atoms is more
limited, though. Anyhow, spatial dependence of the laser field
introduces selective heating of the hot atoms, and
the consequent escape from the lattice. It would probably be too far 
reaching to claim that this effect should be visible
in an experiment. Our model neglects the
reabsorption of photons which may affect 
the total thermodynamics of the atomic cloud
at high densities, and we neglect also
the Doppler cooling. Thus we have revealed one
aspect of the thermodynamics 
of a densely-populated near-resonant optical
lattice but the solution for the complete problem
is simply out of reach for the modern computational
resources.

Figure \ref{fig:Narrowing1} shows an example of the results
for collisions in a red-detuned lattice from Ref.~\cite{Piilo02a}.
This
example is for a sodium lattice of depth $U_0=339E_r$.
The comparison is done between the momentum
probability distributions of the interacting and non-interacting
cases with the occupation density of $25\%$ of the lattice.
The central peak is clearly narrower when the interactions
between the atoms have been included. This central
peak corresponds to the atoms which are trapped
in a lattice, and the wide wings correspond to background atoms
which are presumably out of the recapture range of the lattice
and ejected from the lattice.

One can calculate by semi-classical means the excitation
and survival probability for various molecular potentials.
We have done a simple semi-classical analysis by
using Landau-Zener approach
in Ref.~\cite{Piilo02a}. This analysis supports the conclusions
presented above and shows the high
probability  for the atom pair to gain kinetic
energy by the amount with which the collision
partners are kicked out from the lattice.
 
\subsection{Blue-detuned lattices}

The prospect of using
the trapping and cooling lasers
for efficient optical shielding has been studied in Ref.~\cite{Piilo02b}.
Complete optical shielding would make collisions
between atoms, when they end up in a same lattice site,
elastic, and it would also prevent atoms from close
encounters, reducing, e.g., inelastic 
hyperfine changing collisions. 
Thus efficient optical shielding
could be beneficial in optical lattices
in addition to the
typical
darkness of the blue-detuned lattices. 
The number
of scattered photons in gray-lattices can be roughly
two orders of magnitude smaller than in MOTs
\cite{Hemmerich95a}.
The role of the radiation pressure due to the reabsorption 
of photons diminishes, and our simplified model describes in a more 
realistic way the total thermodynamics of the atomic
cloud, not only the collision aspect of the thermodynamics.

\begin{figure}[t!]
\noindent\centerline{\psfig{width=70mm,file=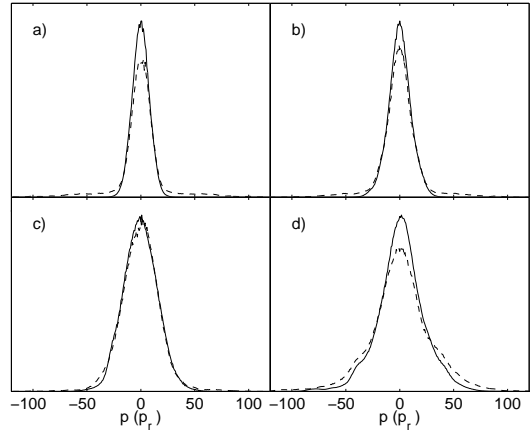} }
\caption[f2]{\label{fig:BlueResults1}
The momentum probability distributions
 for the $\delta=5 \Gamma$ blue lattice case. The Rabi
frequencies are: a) $\Omega=1.5\Gamma$, b) $\Omega=2.0\Gamma$, c)
$\Omega=3.0\Gamma$, and  d) $\Omega=5.0\Gamma$. The momentum
$p$ is expressed in the recoil unit $p_{r}=\hbar k_{r}$. The 
dashed line is for
the interacting and the solid line for the non--interacting
atoms. When $\Omega$
increases the regime changes from incomplete shielding, a) and b),
to complete shielding, c),  and finally to off--resonant heating in d).
The momentum distributions also get wider due to the deepening of the 
 lattice with
increasing $\Omega$. }
\end{figure}

\begin{figure}[t!]
\noindent\centerline{\psfig{width=70mm,file=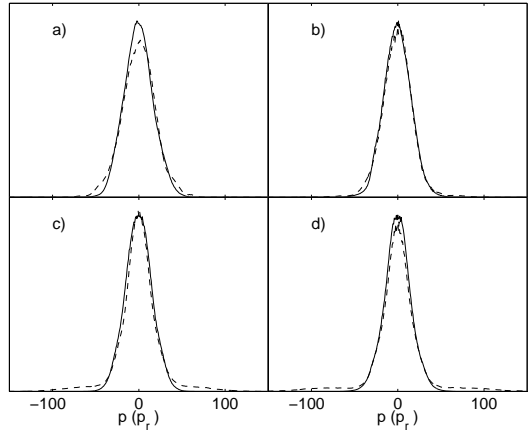} }
\caption[f2]{\label{fig:BlueResults2}
Momentum distributions for fixed $U_{0}\sim 710E_{r}$.
a) $\delta=1.5\Gamma$
b) $\delta=5.0\Gamma$
c) $\delta=7.0\Gamma$
d) $\delta=10.0\Gamma$.
The off-resonant processes play a role at the small detuning, a). In the
intermediate detuning shielding may become complete and the collisions
between atoms are elastic, b). For larger detuning the shielding is
incomplete, c) and d).}
\end{figure}

Figure~\ref{fig:BlueResults1} presents the results of the simulations
where the detuning is fixed to $\delta=5\Gamma$ \cite{Piilo02b}. The results show
clearly how the efficiency of the shielding changes.
When the coupling laser is weak, a large number of the 
collisions are still inelastic ones. Wide wings appear in the momentum
probability distribution, see Figs.~\ref{fig:BlueResults1} (a) and 
\ref{fig:BlueResults1} (b). All the population, which is excited
at the Condon point, does not return to the ground state
when the atoms move apart again. The quasimolecule 
slides down on the tail
of the repulsive state producing a mild heating
effect \cite{Suominen95a}. 
For a moderate field strength it is hard to see differences
between the distributions for the interacting and non-interacting
atoms, Fig.~\ref{fig:BlueResults1} (c).
The optical shielding has become complete and
atoms collide elastically when they end up in 
the same lattice site. The collision partners begin to move
apart at large internuclear distances and also the short
range unwanted effects are avoided. A further increase
in the laser field strength makes it possible for the population
to be excited into the attractive molecular state by 
off-resonant means \cite{Suominen96c}.
This is the reason for
the deviation of the two distributions in 
Fig.~\ref{fig:BlueResults1} (d), where the appearance of
the wings is also qualitatively different than in the weak-field
case of incomplete shielding.

Figure~\ref{fig:BlueResults2} presents another view to
the shielding studies \cite{Piilo02b}. Instead of keeping the laser
detuning fixed,
here the lattice modulation depth is kept nearly constant.
It turns out that for a very small detuning, 
Fig.~\ref{fig:BlueResults2} (a), the off-resonant effects heat
the atomic cloud. This corresponds to the regime
where $\Omega/\delta>1$, and the steady state formation
between the ground and the excited states surpasses the dynamical
resonant excitation process. When the detuning
is increased, the point of complete shielding is reached,
Fig.~\ref{fig:BlueResults2} (b). In the region where
$\Omega/\delta\ll1$, Fig.~\ref{fig:BlueResults2} (c) and (d),
shielding becomes incomplete again due to
the weak excitation and stimulated re-excitation in Condon
point.

The results demonstrate clearly that 
the co-existence of cooling, trapping, and shielding
processes is possible in blue-detuned near-resonant
optical lattices. The shielding is not always complete
but by the careful choice of parameters shielding
becomes very efficient.
Moreover, this can be obtained
within a typical and convenient parameter regime
for near-resonant lattices,
e.g., in Fig.~\ref{fig:BlueResults1} (c) $\delta=5\Gamma$,
$\Omega=3\Gamma$, and $U_0=712E_r$.
A clear advantage here is the achievement
of complete shielding with the same lasers
which provide the cooling and trapping. This is
in contrast to MOTs where one needs to introduce
additional lasers for shielding.

Even though the available occupation densities 
in near-resonant optical lattices have been 
very low so far, the metastable rare-gas atoms
could provide a convenient case for an experimental
study of shielding in optical lattices
due to the clear ion signal that marks collision
events
\cite{Katori94a,Walhout95a}.

The experimental work on optical shielding
in MOTs show the saturation of the shielding phenomena
when the intensity of the laser field is increased \cite{Weiner99a}.
The saturation has not been present in earlier
theoretical studies of shielding \cite{Suominen95a,Suominen96c},
and we do not
see the saturation of shielding here either. 
The results presented in
Ref.~\cite{Piilo02b} seem to confirm the view that the saturation does not
arise due to spontaneous emission 
effects \cite{Suominen95a}.
The reason for the saturation of shielding is still
unclear. It has been attributed to various processes, in addition to the
above-mentioned premature termination of shielding via spontaneous
emission~\cite{Suominen95a}. Other possibilities include counterintuitive or
off-resonant processes involving different partial waves, or other processes
that similarly involve multiple states (in contrast to the basic two-state
approaches~\cite{Weiner99a,Suominen96b,Yurovsky97a,Napolitano97a}).
In the case when several partial waves are considered,
a net of Condon points (instead of only one resonant
point of two-level models) opens a door 
for complicated excitation and de-excitation patterns
which may affect the system dynamics \cite{Piilo04a}. 
The partial wave description is a useful frame of reference
in normal collision studies to allow for more than one dimension.
Due to the low collision velocities one can limit the studies
to a few waves only. But the presence of an optical lattice
forces us to abandon this description. Since our model is, 
due to the computational limitations, only one-dimensional,
we can not conclusively that saturation of shielding should be absent
in a lattice experiment. 

So far there has been
very few cold collision experiments in optical
lattices \cite{Kunugita97a,Lawall98a}.
These experiments showed how the lattice
structure affects the transport of atoms
by using collisions as a probe.
We hope that our work serves as a motivation
for experimentalists to do shielding studies in
blue-detuned optical lattices.

\subsection{Collision rates}

The results of Refs.~\cite{Piilo01a,Piilo02a} 
show that in our selected parameter
regime, i.e., the parameter regime for near-resonant lattices,
the 
motion of the atom
 {\it between} the lattice sites
(or in the lattice site occupied by the single atom only)
is not strongly
affected by other atoms, not at least for
the occupation densities which
we have used (maximum $25\%$). 
Hence, binary interactions between the atoms come into play only when 
two atoms simultaneously occupy the same lattice site.
This makes
it possible to develop a method to calculate
the average rate at which two atoms  end up in the same site, and
the consequent 
cold collision rate in a 
lattice, by following the trajectories
of single atoms  \cite{Piilo03a}. Cold collision rate refers here to the 
average rate of atoms
to reach the region of the resonant Condon point in the presence
of near-resonant light and
describes the rate of occurrence of radiative heating
events in a red-detuned lattice, or optical shielding events,
if blue-detuned light is used. It is
assumed that the two atoms always collide when they end up in the
same lattice site. For the parameters used here, 
this assumption is confirmed
by the results in Refs.~\cite{Piilo01a,Piilo02a}. 

The basic idea of the developed method is as follows. A possible trajectory
of an atom in optical lattice is given by a single MCWF realization.
Trajectories
in position space
can then give information about the rate at which atoms travel
over the average distance $z_a$ between the 
atoms, which
in turn gives information about the binary collision rate
in a lattice. The average distance between atoms $z_a$
corresponds to the mean free path
of atoms between collision events in our one-dimensional
model. Thus, if we monitor the transport of atoms in 
the lattice over the average distance between the
atoms, or the atomic flux over the average distance,
we also obtain information about the
collision rates.

\begin{figure}[t!]
\centering
\scalebox{0.4}{\includegraphics{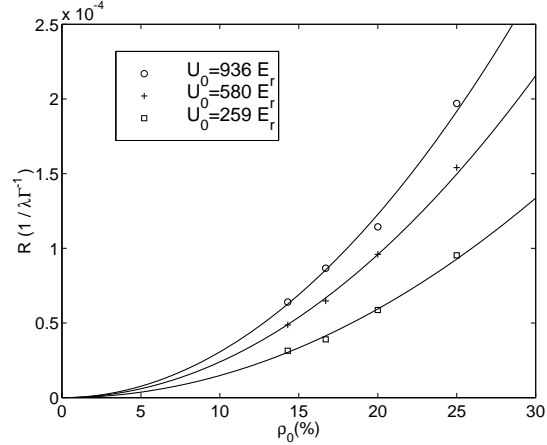}}
\caption[f3]{\label{fig:Crates1}
The binary collision rate $R$
for three different lattice depths $U_0$
as a function of occupation density
$\rho_0$ of the lattice \cite{Piilo03a}. The points show the simulation
results, and the solid lines the quadratic
collision rate curves averaged from the simulation
results for the specific red-detuned lattice. 
The detuning is kept constant at $\delta=-3.0\Gamma$
whereas the Rabi frequency increases
as $\Omega=1.0\Gamma, 1.5\Gamma, 1.9\Gamma$ with the increasing lattice depth indicated in the legend.
}
\end{figure}

Figure \ref{fig:Crates1} displays the calculated collision rates
as a function of the occupation density of the lattice
for three different lattice depths \cite{Piilo03a}. In our simple
1D case the collision rate does not depend on the scattering
cross section and collisions are a measure of transport
in a lattice. 
This was also the case in the experimental
study  of Ref.~\cite{Lawall98a}. In both
cases the collision rate has a quadratic behavior.

The points in  Fig.~\ref{fig:Crates1} are the simulation
results and the solid lines quadratic fits. It is interesting
to note from the Monte Carlo point of view  that we get
results for a wide density range by doing simulations
for only a very few values of density.
The possibility of obtaining
the result for all the values of the variable,
in this case occupation density, from single
Monte Carlo ensemble is a new feature in the MCWF
simulations to our knowledge, at least when
the MCWF method is applied to cold collision problems.

Two-atom collision simulations in a lattice described 
in previous Subsections
are computationally very heavy. It would be useful
to find more simple means to do collision studies in
optical lattices. The method presented above
presents a step in
this direction.
For example, if the semiclassical analysis
shows that for particular parameter values
the colliding
atoms have a high probability to be ejected
from the lattice due to radiative heating, then
the collision rate described here gives directly
the loss rate of atoms from the lattice.
 MCWF simulations for one atom,
like the ones reported in Ref.~\cite{Piilo03a}, are
fairly simple and fast to perform. This
is especially true when compared to the 
two-atom case. Thus the
combination of these simple one-atom simulations
with
semiclassical models for
intra-well collision effects 
have a potential to simplify the studies of binary collisions
in optical lattices.

\section{Conclusions}\label{sec:con}

We have studied the cold collision dynamics between atoms
in near-resonant, dissipative red- and blue-detuned optical lattices.
The applied methods have been mainly based on the Monte Carlo
wave-function method. A semiclassical analysis
has been done which supports the conclusions drawn
from the full quantum-mechanical
calculations \cite{Piilo02a}.

The implementation of the MCWF method to study
cold collisions in optical lattices is not straightforward
and the simulations have been very demanding 
from the computer resource point of view.  This
is due to the internal structure of atoms,
coupling to the electromagnetic environment, 
position dependent
coupling of the atoms to the laser field, and
position dependent coupling between
the atoms.

The results for near-resonant red-detuned lattices are in quite
a sharp contrast to the interaction studies
in magneto-optical traps. Instead of a heating,
a cooling due to the selection
of collision partners from high kinetic energy atoms
is seen in the simulation results.
The blue-lattice results show the applicability
of optical shielding. Future collision
studies require simplifications, for which we propose
a simple way to calculate the collision rate in
optical lattices.

In the past there has been many studies of cold
collisions in magneto-optical traps, see the review \cite{Weiner99a}.
The work presented in here extends the
regime of cold collision studies into the realm
of optical lattices.
This is far from being a trivial step. The major reason 
is that for
sub-Doppler cooling mechanisms, which
exploit various polarization states of a laser field, it is necessary
to account for the internal structure of atoms,
and this greatly complicates the total
system under study and the calculations. 
Moreover, it is not enough to formulate the problem
using only  
the relative
motion between the atoms in a constant laser field.
The position of the atoms with respect to a lattice
structure has to be accounted also.

In conclusion, we have shown how research on
cold collisions in the present of near-resonant light can be extended
from magneto-optical traps to cover also optical lattices.
In future, the recently developed double optical
lattice \cite{Ellmann03a} opens interesting prospects for the experimental
study on cold collisions in dissipative optical lattices.

\ack
The work has been mostly
done at the Helsinki Institute of Physics. 
We acknowledge financial support
from the Finnish National Graduate School
on Modern Optics and Photonics (JP),
the Academy of Finland (projects 206108, 211238, 105740,
and 204777), the Magnus Ehrnrooth Foundation (JP), 
the European Union IHP Network CAUAC,
and thank
the Finnish IT-Center for Science (CSC) for the available
computer resources.
We thank K. Berg-S{\o}rensen for the collaboration
on the collision studies in red-detuned lattices.

\appendix

\section{Required computational resources}

The numerical simulations are demanding since we are dealing with a
36 level quantum
system including various position dependent couplings and a dissipative
coupling to
the environment. We have used 32 processors of an SGI Origin 2000 machine,
which has  128 MIPS
R12000 processors of 1 GB memory per processor
\cite{csc}. The total
memory taken by a
single simulation (fixed $\delta$, $\Omega$, occupation density $\rho_o$,
and atomic species) is 14 GB,
and generating a single history requires 6 hours of CPU time
in red-detuned lattice studies.
 A simulation of 128
ensemble members then requires a total CPU time which is roughly 
equal to one month. The
normal clock time is, of course, much shorter  (roughly 22 hours)
since we take
advantage of powerful parallel  processing for which the
MCWF simulations suit very well.

\end{document}